[1994/12/01]
\documentclass[manuscript]{article}
\title{Dynamic Level Sets}
\author{Michael Stephen Fiske}

\usepackage{amsmath,amsthm,amsfonts,amssymb,mathtools,braket}
\usepackage{graphics}
\usepackage{url}

\chardef\bslash=`\\ 

\newtheorem{thm}{Theorem}[section]

\newtheorem{lemma}[thm]{Lemma}

\theoremstyle{definition}
\newtheorem{defn}{Definition}[section]

\theoremstyle{remark}
\newtheorem{remark}{Remark}[section]

\newcommand{\eval}[2][\right]{\relax
  \ifx#1\right\relax \left.\fi#2#1\rvert}

\begin{document}

\maketitle

\begin{abstract}
A mathematical concept is identified and analyzed that is implicit in the 2012 paper \textit{Turing Incomputable Computation}~\cite{fiske2012}, presented at the Alan Turing Centenary Conference (Turing~100, Manchester). The concept, called \emph{dynamic level sets}, is distinct from classical level sets, the Osher--Sethian level set method, and other standard mathematical concepts in dynamical systems, topology, and computability theory. The paper explains this new mathematical object and why it may have escaped prior characterizations, including the classical result of de~Leeuw, Moore, Shannon, and Shapiro~\cite{deleeuw1956} that probabilistic Turing machines (with bias $p$ where $p$ is Turing computable) compute no more than deterministic ones. A key mechanism underlying the concept is the Principle of Self-Modifiability, whereby the physical realization of an invariant logical level set is reconfigured at each computational step by an incomputable process.
\end{abstract}

\section{Level Sets in Pure Mathematics}

In standard analysis, topology, and dynamical systems theory, a \emph{level set} of a
real-valued function $f : X \to \mathbb{R}$ at value $c$ is the preimage
\[
  f^{-1}(c) \;=\; \{\, x \in X \mid f(x) = c \,\}.
\]
Closely related are the \emph{sublevel sets} $f^{-1}((-\infty, c]) = \{x \mid f(x) \le c\}$.

While contour lines\footnote{Charles Hutton used them in his 1774 survey of Schiehallion
mountain in Scotland.}  on maps \cite{hutton1778} (curves of equal elevation) 
represent the earliest graphical use of the concept, the first \emph{systematic mathematical use of level sets in
dynamics} appears in Lyapunov's 1892 doctoral dissertation~\cite{lyapunov1892}.
Lyapunov's stability analysis depends on the geometry of sublevel sets of a positive-definite
function $V$: the invariance of these sets under the flow verifies stability, while their
outward drift proves instability.

Level sets in classical dynamical systems are static geometric objects. 
The phase space itself, where level sets reside,  was co-developed in the late nineteenth century by
Boltzmann~\cite{boltzmann1872}, Poincar\'e~\cite{poincare1881}, and
Gibbs~\cite{gibbs1902}. Boltzmann's 1872 paper introduced the distribution function on
position-momentum space; Poincar\'e's four-part memoir of 1881--1886 established the
qualitative geometric theory of trajectories in the plane; Gibbs's 1902 treatise
systematized the ensemble formalism. In this framework, the level sets of any function
associated with a dynamical system $\dot{x} = F(x)$ (conserved quantities, Lyapunov
functions, or the vector field itself) are fixed structures. Trajectories move
\emph{through} them, but the level sets do not move.

The entire geometric theory of dynamical systems (invariant manifolds, stable and
unstable manifolds, basins of attraction, separatrices) rests on this assumption, as
covered by the foundational survey of Smale~\cite{smale1967} and in the Russian school's
treatment of bifurcations~\cite{arnold1988,arnold1994}. In the latter, bifurcations alter
the topology of level sets at isolated parameter values, but always according to equations
fixed in advance; between bifurcation points the level set structure is entirely permanent.

\begin{remark}
The permanence of level sets in classical dynamics is not a technical limitation but a
\emph{foundational assumption}. A classical dynamical system is a triple $(T, X, \Phi)$
where $\Phi : X \times T \to X$ is a fixed evolution rule. The level sets of any function
associated with the system are \emph{static} geometric objects: they do not move, deform,
merge, or split. Trajectories move \emph{through} them; the level sets themselves are
permanent structures in phase space.

\medskip 

This is a stricter condition than merely saying 
the level sets are \emph{determined} by $\Phi$. A level
set can be fully determined by a fixed rule and still change in time --- as in the
Osher--Sethian method described below, where the governing PDE is fixed in advance yet the
zero level set $\Gamma(t)$ moves. 
The distinguishing feature of the classical notion of level sets is 
that stasis is a stricter requirement than determinacy: 
having a fixed rule is not sufficient to assure that level sets are permanent geometric structures, 
as the Osher--Sethian method illustrates.
\end{remark}

\subsection*{The Osher--Sethian Level Set Method (1988)}

A conceptually related but distinct notion was introduced by Osher and Sethian~\cite{oshersethian1988}
in the context of interface propagation and computational fluid dynamics.
In their method, one introduces a function $\varphi : \mathbb{R}^n \times [0,\infty) \to \mathbb{R}$
and tracks the zero level set
\[
  \Gamma(t) \;=\; \bigl\{\, x \in \mathbb{R}^n \;\bigm|\; \varphi(x,t) = 0 \,\bigr\}
\]
as a moving interface. The evolution of $\varphi$ is governed by the \emph{fixed} PDE
\[
  \frac{\partial \varphi}{\partial t} +  F | \nabla \varphi | = 0,
\]
where $F$ is the speed that is a function of the curvature $\kappa$ and 
$| \nabla \varphi |$ is the Euclidean norm of the gradient of  $\varphi$.
Note that the defining rule for the zero level set is specified once and for all in advance.

A crucial point is the following. The \emph{rule} governing $\varphi$ is fixed: it is
a predetermined PDE at $t=0$, set before the evolution of the geometric dynamical system starts. 
What changes in time is the \emph{level set} $\Gamma(t)$ itself -- surfaces may merge, split, or otherwise deform, but
always as the deterministic output of the fixed governing equation. In this sense the
Osher--Sethian method preserves the classical assumption that the evolution rule is
permanently fixed; the novelty lies in representing a moving interface implicitly through the
changing zero level set of a fixed-rule function, rather than tracking the interface explicitly.

The  Osher--Sethian level set differs from dynamic level sets defined in Section~\ref{sect:new_object} below, 
where it is the \emph{logical level set} (as a mathematical set) that is invariant, while its \emph{physical
instantiation} carries no fixed representation -- the representation is not determined in advance, but
reconfigured at each step by an incomputable process that uses self-modifiability. 


\section{The Principle of Self-Modifiability}\label{sect:self_modifiability}

A computer program can be viewed as a discrete, autonomous dynamical system~\cite{fiske2020}. 
A set of differential equations specifies a different type of dynamical system 
that can sometimes model analog machines~\cite{maxwell1868}.  When a dynamical system that 
performs a task is changed so that the task is no longer adequately performed, it is 
desirable to have a process or mechanism to heal the system.  This means that 
part of the system should detect a change and part of the system should {\it self-modify}  
the damaged dynamical system back to the original system that adequately performed 
the task.   The capability of a dynamical system to change itself is called {\it self-modifiability}.
This principle was explicitly introduced in~\cite{fiske2023healing}.

A more concrete way to think about self-modifiability is to view a dynamical system as a system that 
is governed by a collection of rules.  In the case of differential equations, each equation is a rule;  
in a computational machine, each machine instruction is a rule.  
A dynamical system is self-modifiable if it can change its own rules:  this means the dynamical 
system can add new rules to itself; it can replace current rules with new rules; 
or it can delete rules.   Generally, a self-modifiable dynamical system is {\it non-autonomous} because the rules 
governing its behavior can change as a function of time.

In \cite{fiske2011aem}, the active element machine (AEM) is a computational machine that can add new rules
when a ``dynamical event" occurs during the machine's execution.   
The AEM uses ``meta commands" to change its own rules.  
More generally, self-modifiability need not always be triggered by damage to a dynamical system:  
in the next section, we discuss how the AEM uses self-modifiability as a routine mechanism, 
reconfiguring the physical realization of level sets at every computational step.
 
\newpage 

\section{Level Sets in the Active Element Machine}

The Active Element Machine (AEM)~\cite{fiske2011aem}, when executing a Universal
Turing Machine (UTM) program $\eta$, physically realizes the boolean functions (representing machine instructions)
\[
  \eta_k : \{0,1\}^3 \times \{0,1\}^2 \;\longrightarrow\; \{0,1\},
  \quad k \in \{0,1,2,3,4,5\},
\]
as active element firing patterns. Each function $\eta_k$ has a fixed logical level set
structure. For example,  $\eta_3^{-1}\{1\} \;=\;$

\medskip 

  $\bigl\{\,(111,00),\, (110,00),\, (110,01),\, (110, 10),\, (101, 00),\, (101, 01),\, (101, 10),$  
       
\hskip 0.7pc  $(100, 00),\, (100, 10),\, (011, 01),\, (011, 10),\, (010, 11),\, (010, 01),\, (010, 00)  \bigr\}$,  

\medskip 

\noindent which is a property of the Minsky UTM program being executed.

What is \emph{not} fixed is the \emph{physical realization} of these level sets within the
AEM. At each UTM computational step, quantum random bits $R_0, R_1, \ldots, R_{13}$
are generated by a quantum random number generator of the type described
in~\cite{stipcevic2007}; see also~\cite{herrerocollantes2017} for a comprehensive review
of such devices. Meta commands \texttt{set\_dynamic\_C} and \texttt{set\_dynamic\_E} then
\emph{reconfigure} the active elements $D_0, \ldots, D_{13}$ and their connections
(the physical objects that compute membership in $\eta_3^{-1}\{1\}$), based on which of the
$R_k$ fired. The same logical level set is realized by a \emph{different}
spatio-temporal firing pattern at each computational step.

More precisely, for each UTM instruction $(q, \alpha) \in Q \times A$, an
invertible boolean function is defined
\[
  B_{(q,\alpha)} : \{0,1\}^{15} \;\longrightarrow\; \{0,1\}^{15}
\]
(shown in Table~4 of~\cite{fiske2012}) that maps the random firing pattern
$(x_0, \ldots, x_{13}, b_3)$ to the firing activity $(y_0, \ldots, y_{13}, p_3)$ of the
elements computing $\eta_3$. The choice of which $B_{(q,\alpha)}$ governs the realization
at step $j$ depends on the current UTM instruction $I_j$ -- that is, on the \emph{dynamical
state of the computation itself}.

\section{A New Mathematical Object}\label{sect:new_object}

The concept is stated more precisely.

\begin{defn}[Dynamic Level Set]\label{defn:dynamic_level_set}
Let $f : X \to \{0,1\}$ be a fixed boolean function with level sets $f^{-1}\{0\}$ and
$f^{-1}\{1\}$. A \emph{dynamic level set decomposition} of $f$ is a family
of invertible boolean functions $\{B_s\}_{s \in S}$, indexed by a state space $S$,
together with an incomputable sequence $\omega : \mathbb{N} \to \{0,1\}^n$ and a
computable function $h : \mathbb{N} \to S$, such that at step $j$ the physical encoding of
$f^{-1}\{1\}$ is determined by $B_{h(j)}(\omega(j))$ rather than by any fixed
representation. The level set as a mathematical set is invariant;
its physical instantiation has no fixed representation.
\end{defn}

\begin{remark}
Definition \ref{defn:dynamic_level_set} captures what is constructed in Procedure~2 ~\cite{fiske2012}.
The logical level sets $\eta_k^{-1}\{1\}$ are invariant properties of the UTM program
$\eta$. The physical realizations carry no fixed representation within the AEM; each representation is
reconfigured at each computational step by a rule that depends on the current UTM instruction (computable via $h$) 
and on quantum random input.  
\end{remark}

\bigskip 

A dynamic level set is novel for the following reasons.

\paragraph{It is not a parametric family.}
In a parametric family $F(x;\, \theta(t))$, the level sets $F^{-1}(c;\, \theta(t))$ sweep
through a predetermined sequence as $\theta$ varies. Here, no predetermined sequence
exists: the quantum randomness assures that the sequence of level set realizations
is itself Turing incomputable, so it cannot have been fixed in advance.

\paragraph{It is not bifurcation theory.}
In bifurcation theory~\cite{arnold1988,arnold1994}, level set topology changes at isolated
parameter values according to fixed equations of motion. In procedure 2 ~\cite{fiske2012} , the
physical realization changes at \emph{every} computational step, not at exceptional points,
and is governed by an incomputable process rather than a smooth family.

\paragraph{It is not a probabilistic Turing machine in the sense of de~Leeuw et al.}
The classical 1956 result shows that a probabilistic Turing machine with bias 
$p$, where $p$ is Turing computable, computes no more than a deterministic one. 
That result applies because the Turing machine
\emph{program}, and therefore its level set structure, does not change in response to 
random input. In the AEM, the Meta command reconfigures the physical realization of the level
sets at each step based on the quantum random input. The program's \emph{logical} content is
invariant; its \emph{physical encoding} is not.  This is the structural distinction
that allows this new model of computation to escape the 1956 result.

\section{Incomputability}

Lemma~4.1 of~\cite{fiske2012} is a fundamental lemma for proving incomputability.  

\begin{lemma}[Fiske 2012]
Let $\varphi : \mathbb{N} \to \{0,1\}$ be an incomputable function. For each $k$ with
$1 \le k \le m$, let $B_k : \{0,1\}^n \to \{0,1\}^n$ be an invertible boolean function,
and let $h : \mathbb{N} \to \{1,\ldots,m\}$ be a computable function. Define
$g : \mathbb{N} \to \{0,1\}$ by
\[
  \bigl(g(jn+1),\, \ldots,\, g((j+1)n)\bigr)
  \;=\;
  B_{h(j)}\bigl(\varphi(jn+1),\, \ldots,\, \varphi((j+1)n)\bigr).
\]
Then $g$ is incomputable.
\end{lemma}

The proof is by contradiction: if $g$ were computable, then since each $B_k$ is invertible
and $h$ is computable, one could recover $\varphi$ by applying $B_{h(j)}^{-1}$ to each
$n$-tuple, contradicting the incomputability of $\varphi$.

The dynamic level set construction generates an
incomputable firing pattern sequence whenever the source of randomness measures 
quantum events. The Turing incomputability of the observable behavior is a
\emph{mathematical consequence} of the new level set concept, not an additional assumption.

\section{Connection to the Principle of Self-Modifiability}

This concept links the 2012 Turing Centenary paper~\cite{fiske2012} and 
subsequent papers on self-modifiable dynamical systems~\cite{fiske2023healing,fiske2024}. 
In the AEM executing the UTM:

\begin{itemize}
  \item The \emph{state space} is fixed: the set of UTM configurations.
  \item The \emph{logical transition function} $\eta$ is fixed: it is a specific UTM
        program.
  \item The \emph{physical realization of the level sets} of $\eta$ -- the actual dynamical
        objects computing membership in $\eta_k^{-1}\{0\}$ and $\eta_k^{-1}\{1\}$ -- are
        self-modifying during execution.
\end{itemize}

This construction illustrates the Principle of Self-Modifiability~\cite{fiske2023healing} applied to
the level set structure of the system. The system modifies the geometric objects that
define how it computes, \emph{while it is computing}. The level sets as logical objects
are invariant; the level sets as physical objects are self-modifiable — their representation 
can change via the meta command and quantum randomness. 
This distinction between the logical and 
physical identity of a mathematical structure is itself a new concept.

\section{Level sets and Phase spaces}

The phase space framework inherited from Boltzmann~\cite{boltzmann1872},
Poincar\'e~\cite{poincare1881}, and Gibbs~\cite{gibbs1902} treats every geometric object
in the phase portrait -- including level sets of the Lyapunov function~\cite{lyapunov1892},
and the stable and unstable manifolds identified in Smale~\cite{smale1967} -- as a permanent
feature of the landscape. The bifurcation theory of the Russian school~\cite{arnold1988,
arnold1994} allows the topology of these objects to change, but only at isolated points
and always according to equations fixed in advance.  None of this allows the 
\emph{physical realization} of level sets to lack a fixed representation.

The closest standard concepts are mathematically distinct. 

\begin{itemize}
  \item \textbf{Non-autonomous systems}: the vector field depends on time, but according
        to a rule fixed before evolution begins.
  \item \textbf{Variable structure control}: the controller switches between
        pre-specified rule sets, but the full menu of rules is fixed before execution.
  \item \textbf{Probabilistic Turing machines}: random input does not change program
                structure, as established in the 1956 result.
\end{itemize}

\noindent  Dynamic level sets, as defined herein, do not appear to have been formalized or studied
in the mathematical literature prior to the paper, ``Turing Incomputable Computation''.

\section{Summary}

The construction in~\cite{fiske2012}, implemented as Procedure 2,  
implicitly introduces a new mathematical object:
a system in which the logical structure of level sets is invariant, but their physical
realization has no fixed representation --
it is reconfigured at each step by the self-modifiability of the meta command in the AEM  
and quantum randomness~\cite{stipcevic2007,herrerocollantes2017}. 
This mathematical object:

\begin{enumerate}
  \item  Escapes the 1956 de~Leeuw et~al. results by using the 
         mechanism of a Meta-command level set realization;

  \item  Generates Turing incomputable behavior from a finite program, as proved by
         Theorem~4.2 via Lemma~4.1 of~\cite{fiske2012};

  \item  Achieves Shannon perfect secrecy~\cite{shannon1949} of machine execution, 
         as proved by Theorem~4.3 of~\cite{fiske2012}; and

  \item  Provides the earliest and perhaps most mathematically explicit 
         instance of the Principle of Self-Modifiability~\cite{fiske2023healing}, 
         implemented with an active element machine~\cite{fiske2011aem}.
\end{enumerate}


\end{document}